\documentclass[twocolumn,amssymb,amsmath,aps]{revtex4}
\usepackage{graphicx}
\usepackage{dcolumn}
\usepackage{bm}
\usepackage{latexsym,epsf,epsfig}
\usepackage{array}
\usepackage{pstricks}
\usepackage{caption}
\usepackage{amstext}
     
\begin{document}

\title{Reply to Comment on "Reappraisal of the Electric Dipole Moment Enhancement Factor for Thallium"}
\author{H. S. Nataraj$^{1}$}
\author{B. K. Sahoo$^2$} \author{B. P. Das$^3$ } \author{D. Mukherjee$^4$}
\affiliation{$^1$ Cyclotron and Radioisotope Center, Tohoku University, 9808578 Sendai, Japan}
\affiliation{$^2$ Theoretical Physics Division, Physical Research Laboratory, Ahmedabad 380009, India}
\affiliation{$^3$ Theoretical Astrophysics Group, Indian Institute of Astrophysics, 560034 Bangalore, India}
\affiliation{$^4$ Indian Association for the Cultivation of Sciences, 700032 Kolkata, India}

\begin{abstract}
In a recent Comment [arXiv:1108.3399], Dzuba and Flambaum have referred to the disagreement of the
results of our latest calculations of Tl electron and scalar-pseudoscalar (S-PS) electric dipole moments (EDMs) and Cs parity non-conservation (PNC) with some other calculations. We have responded to all their points and also discussed the
larger issues related to them. We have attempted to find the reasons for the disagreement between the results of our calculations and those of others. In particular, we have found that the two important reasons for the discrepancies between 
the Tl EDM calculations of Dzuba and Flambaum and ours are the different choice of single particle orbitals and the treatment
of the valence-core correlation effects. We have demonstrated by numerical calculations that the $V^{N-3}$ orbitals
used by Dzuba and Flambaum overestimate the Tl electron and S-PS EDMs at the Dirac-Fock level. The failure of their
suggested consistency test as interpreted by us is explained for systems with strong correlation like Tl.
Also, the importance of understanding the physics underlying different theories on which atomic EDM and PNC calculations 
are based and comparisons between them are emphasized. 
\end{abstract}

\maketitle

 In their Comment \cite{dzuba0}, Dzuba and Flambaum begin by pointing out that our result for
the electron EDM enhancement factor for Tl \cite{nataraj1} is in disagreement with their calculation
\cite{dzuba1} as well as Liu and Kelly's \cite{liu}. They then go on to state with reference to our paper
where this result was reported that "This is one more paper by this group presenting results which
disagree with all other calculations. Others include, e.g. the calculations of the Tl EDM due to 
scalar-pseudoscalar (SPS) CP-odd interaction and parity nonconservation (PNC) in Cs." In this Comment,
they do not, however, make any mention of our Cs EDM calculations based on the relativistic
coupled-cluster (RCC) theory \cite{nataraj2} which agrees well with their calculations \cite{dzuba1}.
With reference to this agreement, Dzuba and Flambaum had stated in their paper \cite{dzuba1} that "Our
calculations are in good agreement with the most recent and accurate calculations". The reference that
they had given for the most recent and accurate calculations was our Cs EDM work \cite{nataraj2}.
Furthermore, they do not refer to our Ba$^+$ and Ra$^+$ PNC papers
\cite{bijaya,wansbeek} in their Comment. In a recent
paper Dzuba and Flambaum \cite{dzuba5} have reported that our Ba$^+$ and Ra$^+$ PNC results are in
agreement with those of their latest calculations.

  We feel it is necessary to put the above remarks of Dzuba and Flambaum on our Tl EDM and Cs
PNC calculations in perspective. A large number of calculations 
of the electron EDM enhancement factor (EDM-EF) for Tl have been performed using a variety of
methods over the past four decades. We have mentioned in our paper \cite{nataraj1} that the results
of these calculations vary from $-179$ to $-1041$. This is also evident from Dzuba and Flambaum's paper
\cite{dzuba1}. The reason for this variation is the peculiarly large electron correlation effects in Tl
EDM-EF. While
it is true that our result does not agree with those of other calculations which were carried out using
approximations substantially different from ours, it is equally true that the result of Dzuba and Flambaum agrees
well with only the calculation of Liu and Kelly \cite{liu}. As we have explained in our paper \cite{nataraj1} this agreement is fortuitous. 
In their paper \cite {dzuba1}, Dzuba and Flambaum simply state this
agreement, without giving any reasons for it. In contrast, we \cite{nataraj1} have endeavoured
to explain the discrepancies between our results and those of Liu and Kelly \cite{liu} and Dzuba
and Flambaum \cite{dzuba1}. For any scientific problem, the mere agreement or disagreement between two results is far less
important than the reasons underlying them. It is therefore imperative to try to understand 
these reasons when one is comparing the results of two different theoretical methods. 
As for the differences between Liu and Kelly's work and ours, we reiterate the following points that
we had stated in our paper \cite{nataraj1}: "The following approximations were made in the former
work (Liu and Kelly) (i) An approximation to only the one electron part of the EDM Hamiltonian is considered,
thereby neglecting the important contributions partly from the DF potential and largely from the two electron
Coulomb interaction. (ii) Only the linear terms and a few
selected nonlinear terms have been used in the calculations. The CC equations have not been fully solved even at
the CCSD level as a coupled-electron pair approximation has been used to solve for the quadratic terms that have
been taken in the unperturbed doubles equation. (iii) A few selected triple excitations are included only in the unperturbed
singles amplitude equations, where as, several dominant triples terms are completely ignored in the unperturbed doubles equations.
Thus, the contribution of triples is taken into account in a nonstandard way. 
(iv) The inner core is frozen up to the 4s orbital for the calculation
of the unperturbed amplitudes, where as, for solving the
perturbed doubles equations, 4s, 4p and 4d orbitals are
further frozen. Such an inconsistent treatment introduces
uncontrollable errors. In contrast to the above drawbacks,
we consider all the nonlinear terms arising from the single
and double excitations. In addition, we consider the leading
triple excitations in both the unperturbed singles
and doubles cluster equations. We solve the unperturbed
and perturbed, closed- and open-shell equations, self consistently
in the framework of the relativistic CCSD(T) approach, taking into account
the excitations from all the core electrons." 

It is evident from the above comparison that Dzuba and Flambaum's
point in their Comment \cite{dzuba0} that Liu and Kelly have used the same relativistic
coupled cluster method that we have used to calculate the EDM-EF for Tl
is extremely specious. It makes very little sense for us to spend a few
months to try and reproduce their result, particularly when we know that
it is based on inconsistent approximations leading to cluster amplitudes that are partially unphysical. Given these  shortcomings of Liu and Kelly's work, 
a better way of testing the
reliability of our Tl EDM-EF calculation would be to compare the results of our 
calculated quantities like allowed electric dipole
transition amplitudes and hyperfine constants that are related to the EDM enhancement factor
with available experimental data. That is precisely what we have done in 
Table IV of our paper \cite{nataraj1}. The overall agreement of our calculations with
measurements is better than that of Dzuba and Flambaum \cite{dzuba1}. Liu and Kelly \cite{liu}
have not performed any of these calculations. We are surprised that Dzuba and Flambaum
have not referred at all to the comparison of the calculations of the relevant Tl transition
amplitude and hyperfine constants in their Comment \cite{dzuba0}. While these properties for
Tl are not as sensitive to electron correlation effects as the EDM-EF, it is necessary
to calculate them in order to get a sense of their accuracies as they are related to the EDM-EF.

With reference to our comments about the comparison between Dzuba and Flambaum
and our calculations in our paper \cite{nataraj1}, we would like to categorically state that they are not
misleading. We had not made any claims about the P and T violating Hamiltonian
that Dzuba and Flambaum had used. Rather, we had surmised the form of their 
Hamiltonian as they had not given any information about it in their
paper. In order to clarify, we reproduce the relevant excerpt from our paper:
"It appears from the previous work of Dzuba and Flambaum
that the P \& T violating Hamiltonian used in [10], considers
only the internal electric field due to the nucleus and not
the electrons; i.e., the entire two body Coulomb potential is neglected."
It is still not clear to us from Dzuba and Flambaum's Comment \cite{dzuba1}, whether they have
included the gradient of the exact or the approximate two-electron Coulomb potential--they
have not given the explicit form of their Hamiltonian.

Indeed, we did mention in our paper that the $V^{N-3}$ core, virtual and valence orbitals of
Dzuba and Flambaum are highly contracted--plots of some of these orbitals are given in Figs.
(a)-(f). We should have worded it more carefully and emphasized that the $V^{N-3}$ valence (6p$_{1/2}$) 
and the virtual (7s) orbitals are highly contracted
at large $r$ relative to their $V^{N-1}$ counterparts as shown in Figs. (d) and (f). The EDM-EF for Tl is 
very sensitive to the choice of 6p$_{1/2}$, 7s and 6s orbitals and it is clearly desirable to use
$V^{N-1}$ orbitals for an accurate determination of this quantity as they exhibit the correct physical 
behaviour, particularly at large $r$ unlike the $V^{N-3}$ orbitals.

Table I highlights the large discrepancies between the $V^{N-1}$ and $V^{N-3}$ Dirac-Fock (DF)
results for Tl EDM-EF and related properties. In retrospect, it appears that this is an 
important reason for the disagreement between the results of Dzuba and Flambaum \cite{dzuba1} 
and our \cite{nataraj1} calculations. The $V^{N-3}$ RCC result for Tl EDM-EF would be $-560$
if one assumes that it changes in the same proportion to the DF as it does in the $V^{N-1}$ case. This
of course is not a rigorous assumption. It would not be out of place to mention here that in view
of our DF results for the two kinds of orbitals, the
agreement between the CI$+$MBPT result of Dzuba and Flambaum \cite{dzuba1} with $V^{N-3}$ orbitals
and that of the linearized RCC calculation of Liu and Kelly \cite{liu} with $V^{N-1}$ orbitals does indeed
appear to be fortuitous as we had remarked earlier based on other considerations.

\begin{figure}[t]
\begin{tabular}{c}
\includegraphics[%
 scale=0.37]{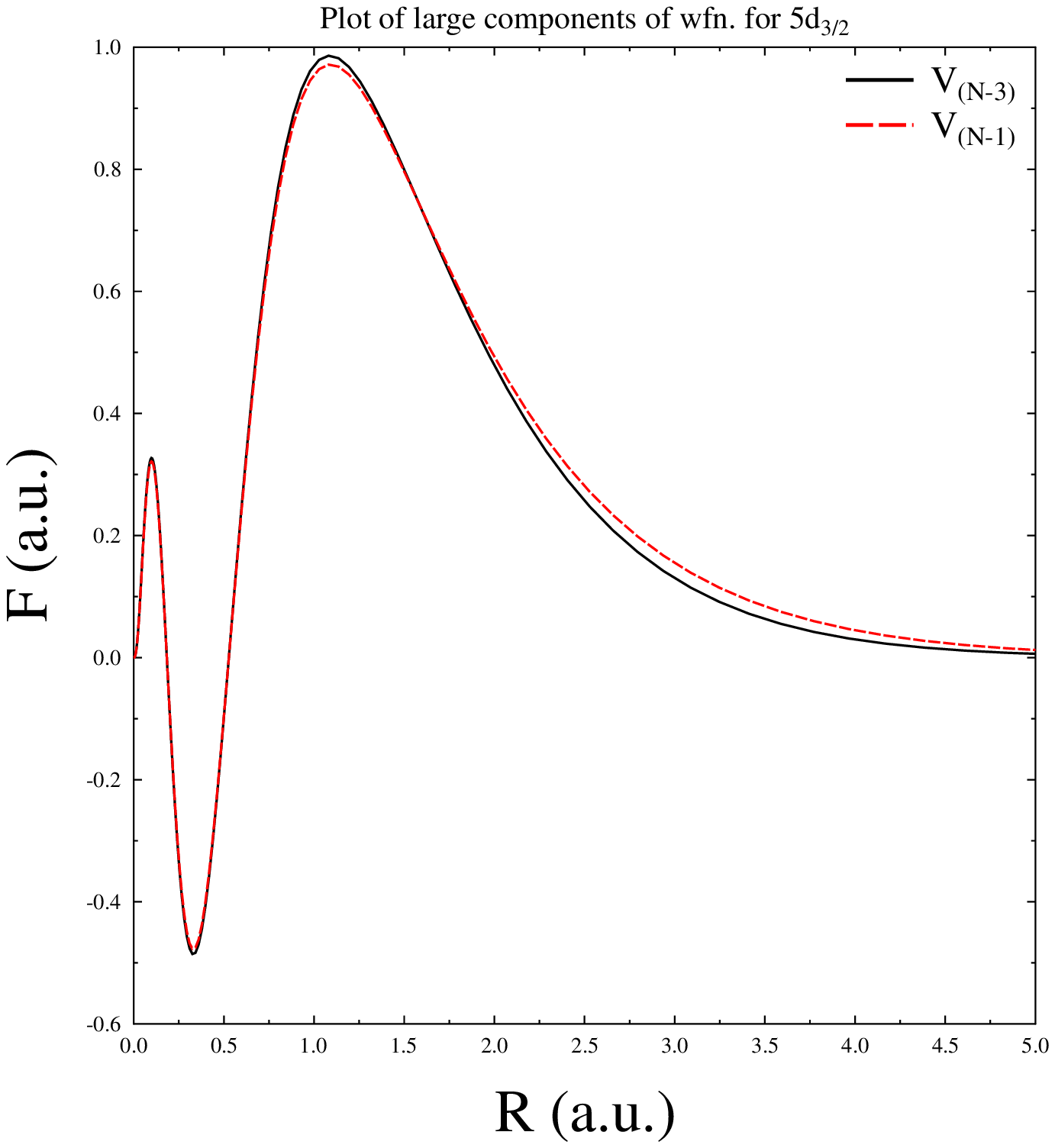}
\tabularnewline
(a) $5d_{3/2}$ orbital wave function
\tabularnewline
\includegraphics[%
 scale=0.37]{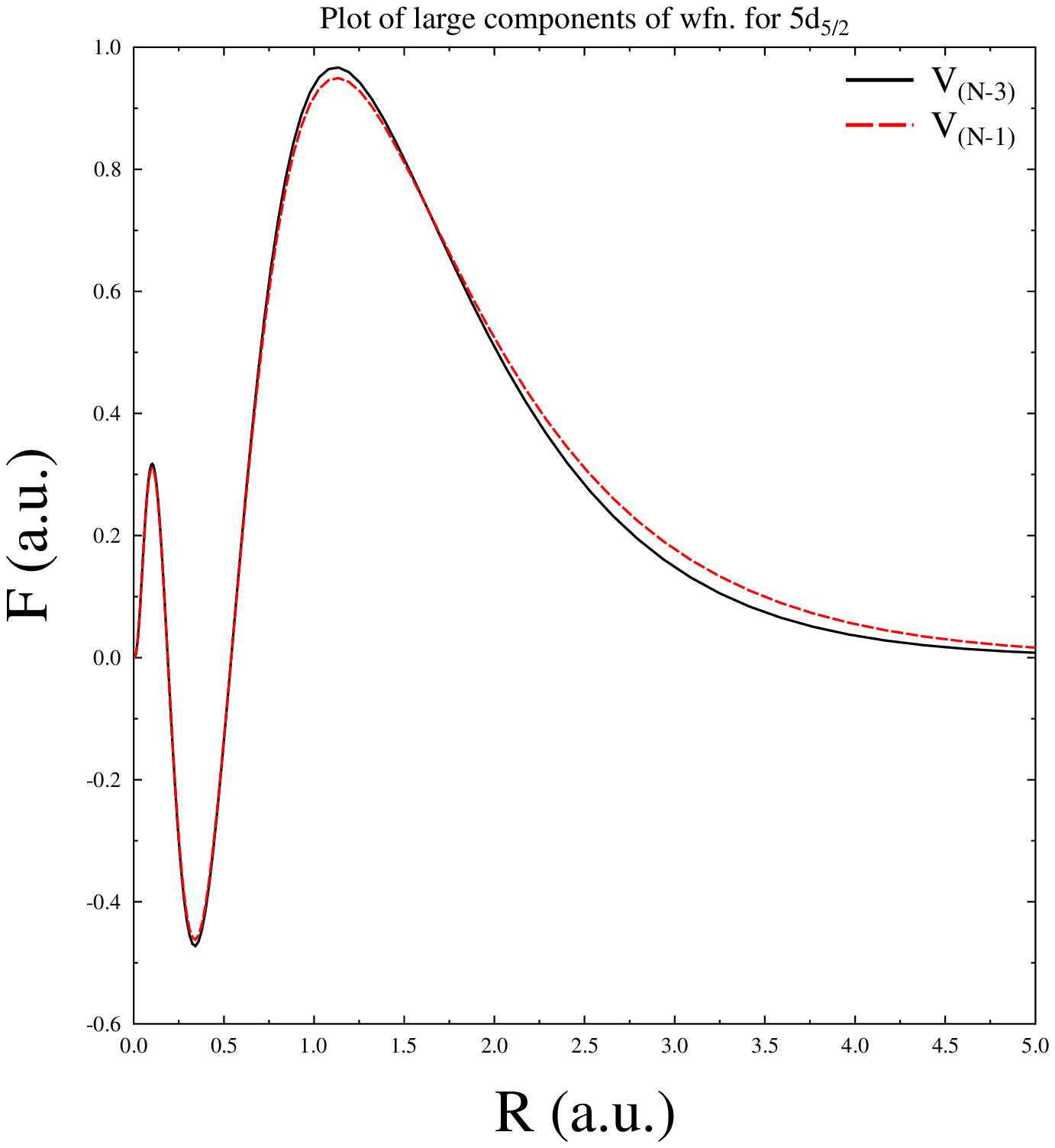}
\tabularnewline
(b) $5d_{5/2}$ orbital wave function
\tabularnewline
\includegraphics[%
 scale=0.37]{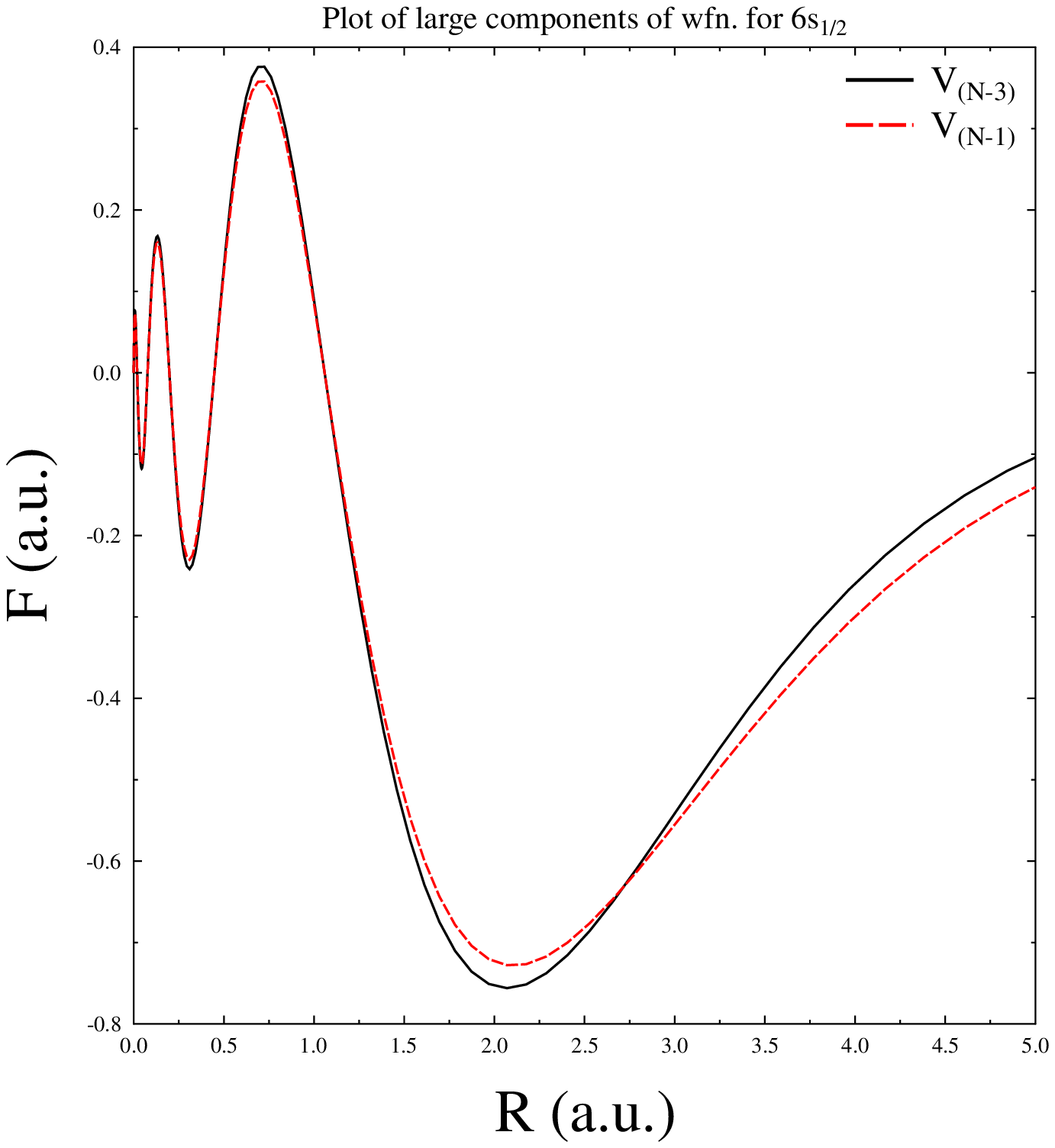}
\tabularnewline
(c) $6s_{1/2}$ orbital wave function
\tabularnewline
\end{tabular}
\caption{Comparison of magnitudes of the $5d_{3/2}$, $5d_{5/2}$, and $6s_{1/2}$ single particle wave functions using $V^{N-3}$ and $V^{N-1}$ potentials.} \label{fig1}
\end{figure}
\begin{figure}[t]
\begin{tabular}{c}
\includegraphics[%
 scale=0.37]{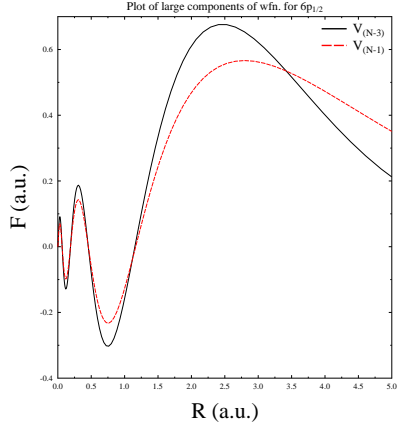}
\tabularnewline
(d) $6p_{1/2}$ orbital wave function
\tabularnewline
\includegraphics[%
 scale=0.37]{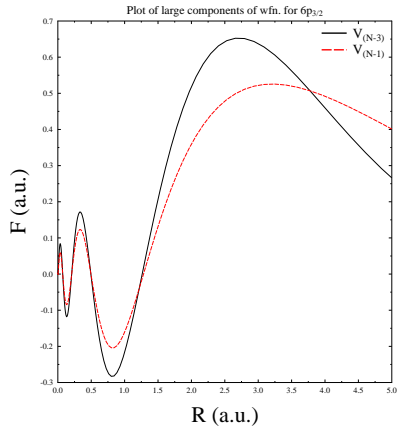}
\tabularnewline
(e) $6p_{3/2}$ orbital wave function
\tabularnewline
\includegraphics[%
 scale=0.37]{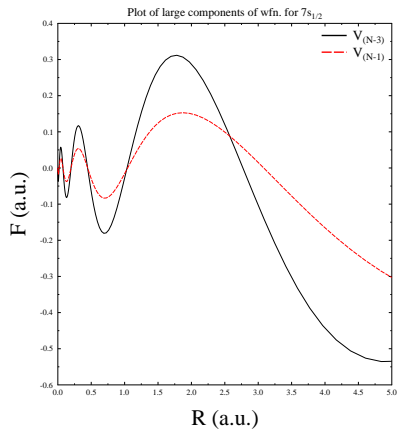}
\tabularnewline
(f) $7s_{1/2}$ orbital wave function
\tabularnewline
\end{tabular}
\caption{Comparison of magnitudes of the $6p_{1/2}$, $6p_{3/2}$, and $7s_{1/2}$ single particle wave functions using $V^{N-3}$ and $V^{N-1}$ potentials.} \label{fig2}
\end{figure}
  
\begin{table}[h]
\begin{center}
 \caption{Various results for different orbitals obtained using $V^{N-3}$ and $V^{N-1}$ potentials in Tl.}\label{tab1}
\begin{tabular}{ccc}
\hline \hline \\
 States/Transitions   & $V^{N-1}$ & $V^{N-3}$ \\
\hline\\
 & \multicolumn{2}{c}{\underline{Enhancement factors due to $d_e$}} \\
 $6p_{1/2}$  & $-422.02$  & $-507.15$ \\ 
 && \\
 & \multicolumn{2}{c}{\underline{S-PS ratio in unit given in \cite{bijaya2}}} \\
 $6p_{1/2}$  & $5.02$  & $6.04$ \\ 
 && \\
 & \multicolumn{2}{c}{\underline{Energy difference in $cm^{-1}$ }} \\
 $6p_{1/2} \ - \ 6s$ & 107507.67  & 60294.73 \\ 
 $6p_{1/2} \ - \ 7s$ & 22714.75  & 67140.58 \\ 
 && \\
 & \multicolumn{2}{c}{\underline{Reduced E1 elements in au}} \\
 $6p_{1/2} \ - \ 6s$ & 2.08  & 2.02 \\ 
 $6p_{1/2} \ - \ 7s$ & 2.05  & 1.26 \\ 
 && \\
 & \multicolumn{2}{c}{\underline{Matrix element of $H_{EDM}^{d_e}$}} \\
 $6p_{1/2} \ - \ 6s$ & 44.26  & 60.89 \\ 
 $6p_{1/2} \ - \ 7s$ & 10.37  & 29.61 \\ 
 && \\
\hline \hline
\end{tabular}
\end{center}
 \end{table}

It is clear from Table I of our paper \cite{nataraj1} that there are delicate
cancellations between various valence-core correlations that we
have evaluated to all orders using the RCC method in the singles, doubles
and partial triples approximation. Dzuba and Flambaum
\cite{dzuba1} have calculated some of these effects (Brueckener pair correlation) mainly by 
second order many-body perturbation theory and have neglected some others like structural
radiation and different classes of higher order RCC terms. The combined contributions from the latter
two effects to Tl EDM-EF are about 30; which is by no means negligible. The approximations
used in \cite{dzuba1} to calculate the valence-core correlations will not be able to capture the 
cancellations in our all order RCC calculations. 

We now turn to the S-PS EDM calculations for Tl. Three relativistic many-body calculations
have been carried out for this quantity \cite{dzuba1, bijaya2,martensson}.
Dzuba and Flambaum's calculation agrees with that of Martensson-Pendrill and Lindroth \cite{martensson}, even
though the approximations used in the calculations are quite different. The latter work considers
one electron effects to all orders, but its treatment of the two electron correlation effects is
rather approximate. It only takes into account the Brueckner pair correlation which is not calculated
directly, but rather it is estimated on the basis of its contribution to the electron EDM-EF of Tl from 
Hartley et al's work \cite{hartley}, which had been performed using the same method that Martensson-Pendrill 
and Lindroth had used in their S-PS EDM calculation \cite{martensson}. Therefore, the latter calculation is 
somewhat less reliable than the former. The S-PS EDM calculations of Dzuba and Flambaum and
Martensson-Pendrill and Lindroth agree, but the electron EDM-EF calculation of the former
is more than three times larger than that of the latter. This is indeed mind boggling and therefore one
cannot attach any importance to the agreement of the Tl S-PS EDM results of the two sets of authors.
Dzuba and Flambaum have not given any explanation for this paradoxical situation. We had employed the RCC
method in the singles, doubles and partial triples approximation \cite{bijaya2} and it is
therefore not at all surprising that our calculation does not agree with the other two calculations which are
based on different approximations.

Dzuba and Flambaum \cite{dzuba0} begin the last paragraph of their Comment by asserting that our Tl EDM-EF \cite{nataraj1}
and S-PS EDM \cite{bijaya2} calculations do not satisfy a simple consistency test. However, they conclude by 
making a rather tentative statement that these two calculations may be internally inconsistent. The consistency test 
in their own words \cite{dzuba0} is as follows: "the ratio of the EDMs due to two operators must be approximately
equal to the ratio of the s−p single-electron matrix elements of these operators. This is because only short 
distances, where single-electron energies can be neglected, contribute to the single-electron matrix elements of 
the CP-odd operators." We do not fully understand
what this test is supposed to mean. Which s-p single electron matrix elements are they referring to?
6s-6p1/2 or 7s-6p1/2  or something else? They have not been very specific. Perhaps the meaning would have been 
clearer if they had supplemented what they have written by mathematical expressions. Towards the end of their 
Comment \cite{dzuba0}, Dzuba
and Flambaum seem to suggest that our value for the ratio they have mentioned in their consistency test is
115 $d_e/C^{SP} \ 10^{−18} \ e cm$ and not 89 $d_e/C^{SP} \ 10^{−18} \ e cm$, which they probably think is the correct value. 
The former is the value of the ratio of the electron EDM-EF to the S-PS EDM for Tl that we have got at
the RCC level \cite{nataraj1,bijaya2} and our value for the same
ratio at the DF level is 84 $d_e/C^{SP} \ 10^{−18} \ e cm$. It therefore appears that the consistency test
that they have referred to in the context of our calculations means our ratios of the values
of the two EDMs for the RCC and DF cases should be approximately equal. In other words, the ratio
of the of electron EDM-EFs of the RCC and DF calculations is approximately equal to the the ratios
of the atomic EDM to the S-PS constant for the same two methods. If our interpretation is
correct then it is straightforward to show that the test that they have proposed does not hold
in general, and in particular, we shall explain why it fails for Tl. In order to do so, we begin
with the unperturbed and first order perturbed wave functions in the RCC approach which can be
expressed as \cite{nataraj1,bijaya2}
\begin{eqnarray}
\vert \Psi_v^{(0)} \rangle &=& e^{T^{(0)}}\{1 + S_v^{(0)} \} \vert\Phi_v\rangle, \ \ \ \text{and}\\ 
\vert \Psi_v^{(1)} \rangle &=&  e^{T^{(0)}}\{ T^{(1)} \left (1 + S_v^{(0)} \right ) + S_v^{(1)} \} \vert\Phi_v\rangle
\end{eqnarray}
respectively, and the electron EDM enhancement factor or in general the ratio of the atomic EDM to a CP violating coupling constant is given by the exact expression 
\begin{eqnarray}
{\cal R} &=& \langle \Psi_v^{(0)}| D |\Psi_v^{(1)} \rangle + \langle \Psi_v^{(1)}| D |\Psi_v^{(0)} \rangle \nonumber \\
 &=& [ \langle \Phi_v \vert \overline{D}^{(0)} \, T^{(1)} + \overline{D}^{(0)} \, S_v^{(1)} + \overline{D}^{(0)} \, T^{(1)} S_v^{(0)} \nonumber \\ && + S_{v}^{(0)\dagger} \overline{D}^{(0)} \, T^{(1)} + S_{v}^{(0)\dagger} \overline{D}^{(0)} \, S_v^{(1)} \nonumber \\ && + S_{v}^{(0)\dagger} \overline{D}^{(0)} \, T^{(1)} S_v^{(0)} \vert \Phi_v \rangle ] +hc \nonumber\\
 &=& {\cal R}_1 + {\cal R}_2 + {\cal R}_3 + {\cal R}_4 + {\cal R}_5 + {\cal R}_6,
\end{eqnarray}
with $\overline{D}^{(0)}=e^{T^{(0)\dagger}} D\, e^{T^{(0)}}$ and each term along with its hermitian conjugate
($hc$) term is given with subscripts from 1 to 6 with their sequence.
To define the single and double excitations, we use the subscripts 1 and 2 for all the RCC 
operators (for the detail see \cite{nataraj1, bijaya2}). As mentioned in \cite{nataraj1, 
bijaya1}, the core and virtual contributions at the DF level comes at the lowest order through
the RCC terms $\langle \Phi_v \vert \overline{D}^{(0)} \, T^{(1)} \vert \Phi_v\rangle$ and
$\langle \Phi_v \vert \overline{D}^{(0)} \, S_{v}^{(1)} \vert \Phi_v \rangle$, respectively.
Denoting ${\cal R}_1= {\cal R}_c^{DF}+{\cal R}_1^{corr}$ and  ${\cal R}_2= {\cal R}_v^{DF}+{\cal R}_2^{corr}$, the total DF result is given by ${\cal R}_{DF}={\cal R}_c^{DF}+{\cal R}_v^{DF}$. With these notations, the above expression can be rewritten as 
\begin{eqnarray}
\frac {{\cal R}}{{\cal R}_{DF}} &=& 1 + \frac {{\cal R}_1^{corr}}{{\cal R}_{DF}} + \frac {{\cal R}_2^{corr}}{{\cal R}_{DF}} + \frac {{\cal R}_3}{{\cal R}_{DF}} + \frac {{\cal R}_4}{{\cal R}_{DF}} \nonumber \\ && + \frac {{\cal R}_5}{{\cal R}_{DF}} + \frac {{\cal R}_6}{{\cal R}_{DF}} .
\end{eqnarray}

All the terms on the right hand side of the above equation are built out of different combinations of
${D}$, ${T^{(0)}}$, ${ S_{v}^{(0)}}$, ${T^{(1)}}$, and ${ S_{v}^{(1)}}$. The last two operators; i.e. ${T^{(1)}}$ and ${ S_{v}^{(1)}}$ contain one order
of the EDM interaction and all orders in the residual Coulomb interaction. The electron EDM interaction Hamiltonian is given by
\begin{eqnarray}
H_{EDM}^{e}&=&2icd_{e}\sum_{j}\beta_{j}\gamma_{j}^{5}p_{j}^{ 2}
\label{eqn1}
\end{eqnarray}
and the S-PS interaction Hamiltonian is given by
\begin{eqnarray}
H_{EDM}^{S-PS}&=&\frac{iG_{F}}{\sqrt{2}}C_{S}\sum_{j}\beta_{j}\gamma_{j}^{5}\rho^j_{nuc}(r),
\label{eqn2}
\end{eqnarray}
where $d_{e}$ is the intrinsic $e$-EDM,
$\gamma^{5}$ is a pseudo-scalar Dirac matrix, $C_{S}$ is the dimensionless
S-PS constant and $\rho^j_{nuc}(r)$ is the $j$th electron density over
the nucleus. The corresponding single particle 
electron EDM and the S-PS matrix elements are given by
\begin{eqnarray}
\langle \phi_i || \text{h}_{EDM}^{e} || \phi_j \rangle &=& - 2c 
 \sqrt{2j_i+1} \delta(\kappa_i,-\kappa_j) \ \ \ \ \ \ \ \nonumber \\
&& \int_0^{\infty} dr ( \frac{{\tilde l}_j ({\tilde l}_j +1)}{r^2} P_i(r) Q_j(r) \nonumber \\ && + \frac{l_j (l_j+1)}{r^2} Q_i(r) P_j(r) \nonumber \\
&& + \frac{dP_i}{dr} \frac{dQ_j}{dr} + \frac{dQ_i}{dr} \frac{dP_j}{dr} )
\end{eqnarray}
and
\begin{eqnarray}
\langle \phi_i || \text{h}_{EDM}^{S-PS} || \phi_j \rangle &=& \frac {G_F}{\sqrt{2}} C_S
 \sqrt{2j_i+1} \delta(\kappa_i,-\kappa_j) \ \ \ \ \ \ \ \nonumber \\
&& \int_0^{\infty} dr (P_i(r) Q_j(r) \nonumber \\ && + Q_i(r) P_j(r) ) \rho^j_{nuc}(r) ,
\end{eqnarray}
respectively, where $l$ and ${\tilde l}$ are the orbital quantum numbers for the large and
small components of the Dirac orbitals.
It is clear that the above two matrix elements are numerically quite different, even though 
both of them are large in the nuclear region.
The amplitudes for ${T^{(1)}}$, and ${ S_{v}^{(1)}}$ are therefore different for the two 
EDMs; particularly in Tl where the electron correlation effects are strong. This is because 
these amplitudes represent different processes involving the EDM and the correlation effects. 
Therefore $\frac{{\cal R}}{{\cal R}_{DF}}$ would in general be different for
the electron EDM and the S-PS interaction making the consistency test that Dzuba and Flaumbaum 
have suggested invalid for
systems with large correlation effects; although it could hold in situations when the correlation
effects are weak or moderately strong. This is reflected in the results of our RCC calculations 
of the two EDMs for Tl--see
the relevant tables \cite{nataraj1,bijaya1}. The trends exhibited by some of the RCC terms in 
the two cases are different--the most striking being the relative contributions of $DT^{(1)}$ 
and $DS_{v}^{(1)}$. As a result, the relative contributions of $\frac{{\cal R}_1^{corr}}{{\cal R}_{DF}}$ and $\frac{{\cal R}_2^{corr}}{{\cal R}_{DF}}$ are quite different.

We would now like to clarify certain issues related to our parity nonconservation (PNC) 
calculation in Cs that the Dzuba and Flambaum \cite{dzuba0} have commented on. It is certainly
not true that our Cs PNC result does not agree with the results of other calculations
of this quantity. In an earlier work, we had reported our Cs PNC result as $0.902(4) \times 
10^{-11} iea_0{-Q_W/N}$ from a preliminary calculation \cite{das} and more recently we obtained
$0.8892\times 10^{-11} iea_0{-Q_W/N}$ (triple excitations involving the valence and two core
electrons were incorporated in the latest result) \cite{bijaya1} using the RCC method. The latter
result is in agreement with another recent calculation by Porsev et al \cite{porsev}, who have got
$0.8891\times 10^{-11} iea_0{-Q_W/N}$ in the framework of the Dirac-Coulomb approximation
using unscaled wave functions by a sum-over-states relativistic coupled-cluster (CC) method at 
the singles, doubles and leading order triple excitations which is at par with our approach.
However, Porsev et al have minimized the uncertainty in the calculation by scaling the wave
functions and they have added other corrections to get the final result. They have used
their final result to probe new physics. In our work, we have only demonstrated the important role
of correlation effects using an {\it ab initio} RCC method and we have not put any effort to
minimise the uncertainty at the Dirac-Coulomb level or considered higher order relativistic and 
nuclear corrections to probe new physics using our results. Our result is not in serious conflict
with any other calculation if it is viewed in the right context. The reported results of 
Dzuba and Flambaum are $0.9001 \times 
10^{-11} iea_0{-Q_W/N}$ \cite{dzuba2} and $0.9078 \times 10^{-11} iea_0{-Q_W/N}$ \cite{dzuba3}
at the Dirac-Coulomb approximation. The same authors had also reported the
s-d PNC amplitude calculations in Ba$^+$ and Ra$^+$ \cite{dzuba4} about a decade ago
\cite{dzuba4}. They had obtained 2.17 and 2.37 (in $\times 10^{-11} 
iea_0{-Q_W/N}$) for Ba$^+$ and 42.9 and 45.9 (in $\times 10^{-11} iea_0{-Q_W/N}$) for
Ra$^+$ using a mixed state and the sum-over-states approaches, respectively. In their Comment,
the same authors had expressed the view that it is necessary to obtain accurate results
for these calculations in order to use them for inferring new physics \cite{dzuba0}. But their
dual results could not have been useful for this purpose. Accurate PNC results for these
ions were only identified when we employed the RCC method and reported our results as
2.46(2) and 46.4(1.4) (in $\times 10^{-11} iea_0{-Q_W/N}$) for Ba$^+$ \cite{bijaya} and 
Ra$^+$ \cite{wansbeek}, respectively. We learnt from a recent work of Dzuba and Flambaum that
using one of their advanced methods, they were able to improve their results to
0.29 and 3.4 (in $\times 10^{-12} iea_0{-Q_W}$) \cite{dzuba1} compared with our results
0.304 and 3.33 (in $\times 10^{-12} iea_0{-Q_W}$) in Ba$^+$ and Ra$^+$,
respectively. We quote below from their paper \cite{dzuba5}: 
"There is also good agreement with Sahoo {\it et al.} for Ba$^+$ [21] and with
Wansbeek {\it et al.} for Ra$^+$ [17] ....".

In conclusion, we have responded to all the points raised by Dzuba and Flambaum in their 
Comment \cite{dzuba0}
on our Tl EDM and Cs PNC calculations. The reasons for the discrepancies between Liu and Kelly 
\cite{liu} and our calculations which were discussed in
our paper \cite{nataraj1} have been reiterated. The two main reasons for the disagreement between Dzuba and Flaumbaum
and our Tl EDM calculations have been identified as the choice of orbitals ($V^{N-3}$ by Dzuba and Flambaum and
$V^{N-1}$ by us) and the treatment of valence-core correlation. 

It has been pointed out that the the agreement of the Tl electron EDM results of Liu and Kelly and Dzuba
and Flambaum is fortuitous. We have provided strong evidence to show that the agreement of the Tl S-PS
EDM results of Dzuba and Flambaum \cite{dzuba1} and Martensson-Pendrill and Lindroth \cite{martensson} raises many more 
questions than it answers. We have argued that the consistency test that Dzuba and Flambaum have referred
to in their Comment involving the electron and S-PS EDMs as interpreted by us is flawed. Our RCC results for quantities
that are associated with Tl EDM are in better agreement with experimental data than those of Dzuba and Flambaum, which
these two authors do not mention in their Comment. Contrary to the comments made by Dzuba and Flambaum on our Cs PNC work, our result does agree with that of another calculation that was performed at the same level of approximation \cite{porsev}.

Dzuba and Flambaum refer to three calculations (Tl electron EDM, Tl S-PS EDM and Cs PNC) in their Comment
for which their results are different from ours. However, in the same Comment, they do not mention the results of our
Cs electron EDM, Cs S-PS EDM, Ba$^+$ PNC and Ra$^+$ PNC calculations which are in agreement with theirs. When the
result of a calculation of ours agrees with theirs as in the case of Cs EDM \cite{nataraj2}, it is described
as accurate by them \cite{dzuba1}, but on the other hand if it does not agree then they make adverse
remarks against it as exemplified in their Comment \cite{dzuba0}. Dzuba and Flambaum have not made any attempt to explain
why a particular result agrees or disagrees with their result \cite{dzuba1,dzuba0}-- they have merely stated
whether it agrees with their result or not. Such an approach provides no insights into the reasons
for the agreement or disagreement between the two methods that were used to obtain these results. An understanding
of the physics embodied by the methods used in the calculations of atomic EDMs and PNC and comparisons between
them are prerequisites for making further progress in this field.

\end{document}